\documentclass[aps,prd,reprint,groupedaddress,showpacs]{revtex4-1}

\usepackage{bm}
\usepackage{graphicx}
\usepackage{hyperref}

\begin{document}


\title{Relativistic Analogue of the Newtonian Fluid Energy Equation with Nucleosynthesis}
\thanks{This manuscript has been authored by UT-Battelle, LLC, under contract DE-AC05-00OR22725 with the US Department of Energy (DOE). The US government retains and the publisher, by accepting the article for publication, acknowledges that the US government retains a nonexclusive, paid-up, irrevocable, worldwide license to publish or reproduce the published form of this manuscript, or allow others to do so, for US government purposes. DOE will provide public access to these results of federally sponsored research in accordance with the DOE Public Access Plan (http://energy.gov/downloads/doe-public-access-plan).}


\author{Christian Y. Cardall}
\email[]{cardallcy@ornl.gov}
\affiliation{Physics Division, Oak Ridge National Laboratory, Oak Ridge, TN 37831-6354, USA}
\altaffiliation[Also at ]{Department of Physics and Astronomy, University of Tennessee, Knoxville, TN 37996-1200, USA}


\date{\today}

\begin{abstract}
In Newtonian fluid dynamics simulations in which composition has been tracked by a nuclear reaction network, energy generation due to composition changes has generally been handled as a separate source term in the energy equation.
A relativistic equation in conservative form for total fluid energy, obtained from the spacetime divergence of the stress-energy tensor, in principle encompasses such energy generation; but it is not explicitly manifest.
An alternative relativistic energy equation in conservative form---in which the nuclear energy generation appears explicitly, and that reduces directly to the Newtonian internal + kinetic energy in the appropriate limit---emerges naturally and self-consistently from the difference of the equation for total fluid energy and the equation for baryon number conservation multiplied by the average baryon mass $m$, when $m$ is expressed in terms of contributions from the nuclear species in the fluid, and allowed to be mutable.
\end{abstract}


\maketitle


\section{\label{sec:Introduction} Introduction}

Most astrophysical environments in which nucleosynthesis occurs are highly dynamic, requiring numerical treatment of at least fluid dynamics and gravity (if not magnetic fields, neutrino transport, etc.). 
One possible approach to studying the nucleosynthesis in such systems (e.g. \cite{Harris2017Implications-fo}) is to begin with simulations that employ a simplified treatment of the nuclear composition, one that is relatively inexpensive computationally but that is sufficiently accurate in terms of the feedback on the fluid from nuclear reactions to get the thermodynamic conditions basically correct.
One important simplification is the exploitation, where applicable, of nuclear statistical equilibrium (NSE; thermal equilibrium, and chemical equilibrium with respect to strong and electromagnetic interactions).
A first approximation at lower densities and temperatures where NSE does not apply is the instantaneous `flashing,' or `flash burning,' of a representative dominant species (e.g. `oxygen' to `silicon,' or `silicon' to NSE).
A better approximation is time-integration of a small `$\alpha$ network' restricted to several species most relevant to energy generation---those between $^4$He and $^{56}$Ni or $^{60}$Zn that are `multiples' of $\alpha$ particles, connected mostly by $(\alpha,\gamma)$ reactions.
Inclusion of Lagrangian `tracer particles' in such simplified simulations then allows the histories of density and temperature of an ensemble of mass elements to be `post-processed' with a larger nuclear reaction network containing many more species of observational interest.
Alternatively, and ideally, a large network could be used within the original simulation in order that local energy feedbacks be more accurately computed.
(An $\alpha$ network also lacks weak interactions, which additionally induce electron fraction feedbacks and energy loss to escaping neutrinos.)

With its more noticeable effect on mass, energy release due to nuclear reactions may be considered an inherently relativistic effect; but
just as the internal energy of a conventional fluid mixture includes the latent heat of phase transitions due to intermolecular forces, and/or the energy of chemical bonds due to interatomic forces, so also a fluid consisting of reacting nuclear species can be handled in an otherwise Newtonian system by including in the internal energy the binding energy of nuclei due to the forces between nucleons.
In principle, therefore, an equation of state can address the physics of nuclear energy generation in a self-contained way, simply by including nuclear binding energy in the definition of internal energy. 

Including the nuclear binding energy in the definition of internal energy is presumably the most natural approach when the matter is assumed to be in NSE, or when a `flashing' approximation is employed. 
In both of these cases, nuclear composition is found not from time integration of reaction rates, but from the assumption of chemical equilibrium with respect to strong and electromagnetic interactions (in addition to thermal equilibrium and charge neutrality). 
This translates into a composition that depends only on instantaneous conditions of density, temperature, and electron fraction---whether rigorously, via the Saha equations in the case of NSE; or phenomenologically, by prescribed thresholds for `flashing.'
A different composition corresponds to a different binding energy; and the instantaneous correspondence of this with a different temperature is naturally expressed in terms of an internal energy function that takes binding energy into account, whose only necessary arguments are density, temperature, and electron fraction. 

In principle, inclusion of nuclear binding energy in the internal energy is also a possible approach when a time-dependent nuclear reaction network is used to track the abundances of species that are not in chemical equilibrium.
The network and fluid equations (including continuity equations for the various species) are numerically advanced in time, yielding at each time step updated internal energy and nuclear abundances; and once again, if the internal energy includes the binding energy of the various species, inversion to find the corresponding temperature (and/or entropy) will reflect the exothermic or endothermic physics of nuclear burning or dissociation.
The complication is that, absent chemical equilibrium (thermal equilibrium and charge neutrality are still assumed), the internal energy function would depend not only on density, temperature, and electron fraction, but on the abundances of all present nuclear species.

In practice, however, the physics of nuclear energy generation has historically been treated separately from the equation of state when a nuclear reaction network has been employed, handled instead as a source term in an energy equation in which the internal energy does \textit{not} include nuclear binding energy \cite{Arnett1996Supernovae-and-} (cf. e.g. \cite{Fryxell2000FLASH:-An-Adapt}).
Presumably the primary reason for this is that a formal decoupling of the physics of nuclear energy generation allows networks of varying size and complexity to be employed without requiring changes to a straightforward equation of state with a simple composition dependence, enhancing code flexibility (and facilitating operator splitting \footnote{Fluid dynamics and nuclear burning/dissociation have distinct physics and consequently different numerical imperatives, particularly when the fluid dynamics is treated in two or three spatial dimensions.
The fluid dynamics can be numerically evolved in a time-explicit manner with one solver, and the nuclear network advanced with a different, time-implicit solver, with updates appropriately combined to advance the overall system forward in time.
A formal separation of nuclear energy generation facilitates the technical implementation of such `operator splitting' by simplifying the composition dependence of the equation of state.}).
For instance, most fluid dynamical studies involving in-situ nuclear reaction networks have assumed conditions in which the nuclei can be treated (aside from reactions) as a nonrelativistic ideal gas mixture. 
That is, when nuclear forces can be ignored between the collisions that induce nuclear reactions, the degrees of freedom associated with such forces can be lumped into a readily separable binding energy contribution, leaving only the translational degrees of freedom as in an ideal gas.
With binding energies thus excluded from the internal energy,
the composition dependence of the baryonic portion of the equation of state is only on the average nucleon number $\bar A$ of the nuclear ensemble (which gives the total particle number density $\sum_a n_a$ of all nuclear species $a$); and the composition dependence of the electron/positron portion of the equation of state adds only a need for the average proton number $\bar Z$ of the nuclear ensemble (in order to specify the total charge density $\sum_a Z_a n_a$ of all nuclear species $a$).



In Newtonian fluid dynamics, nuclear energy generation as a separate source term arises by splitting off the nuclear binding energy from the rest of the internal energy in the first law of thermodynamics applied to a perfect fluid mass element.
When the first law is used in the derivation of an equation in conservative form \footnote{With a source term, this is more properly a `balance equation.' 
But the label `conservative form' is still useful when the sources do not contain derivatives of the fluid variables, to reflect the fact that conservative numerical schemes can still handle shocks without artificial viscosity.} for Newtonian internal + kinetic energy, the nuclear energy generation tags along as a separate source term.

The relativistic approach---which ought to be adopted in the simulation of core-collapse supernovae \footnote{There is at least one example in the literature in which use of full general relativity led to an explosion, while use of Newtonian and `pseudo-Newtonian' gravity produced duds \cite{Muller2012A-New-Multi-dim}. In contrast to this potential qualitative impact of full relativity, the degree of inexact energy conservation arising from numerical errors in typical core-collapse supernova codes is unlikely to make a determinative difference vis-\`a-vis explosions; see Sec.~A.1 of Ref.~\cite{Muller2010A-New-Multi-dim}.} as well as of compact-object mergers---allows a different perspective.
An energy equation in conservative form is given directly by the vanishing spacetime divergence of the stress-energy tensor for a perfect fluid.
This stress-energy tensor contains from the outset not only the internal and kinetic energy densities, but also the rest mass density; hence in principle it already encompasses energy changes due to nuclear reactions. 
Instead of regarding nuclear binding energy as part of the internal energy, it can be included in a rest mass density expressed in terms of an average baryon mass $m$ that varies in time and space.
Teasing the proper time derivative of $m$ out of the formalism is then a conceptually appealing way to isolate nuclear energy generation as a separate source term.

Relativistic numerical fluid dynamics is often prosecuted in conjunction with numerical relativity in some version of the $3+1$ framework. 
This enterprise is sufficiently well developed to be the subject of reviews and books, either with a focus on the fluid dynamics \cite{Wilson2003Relativistic-Nu,Font2008Numerical-Hydro,Rezzolla2013Relativistic-Hy}, or as part of a broader relativistic treatment \cite{Alcubierre2008Introduction-to,Bona2009Elements-of-Num,Baumgarte2010Numerical-Relat,Gourgoulhon201231-Formalism-in,Shibata2016Numerical-Relat}.
But perusal of these works and the literature they point to does not reveal any accounts of nuclear energy generation in fluid dynamics as a direct consequence of a mutable mean baryon mass $m$.

The purpose of this paper is to fill this small but interesting lacuna with a derivation of a fully relativistic fluid energy equation in conservative form that explicitly separates the nuclear energy generation as a source term.
Also of interest is that the resulting evolved energy does not include baryon rest mass, reducing to the `Newtonian total energy' (internal + kinetic) in the appropriate limit.
After reviewing the Newtonian formulation in Sec.~\ref{sec:Newtonian}, this relativistic derivation is presented in Sec.~\ref{sec:Relativistic}, followed by conclusions in Sec.~\ref{sec:Conclusion}.
An Appendix comments on the case in which nuclear binding energy is included in the internal energy, so that there is no nuclear energy generation source term.
A perfect fluid with vanishing viscosity, vanishing heat flux, and local thermal (but not chemical) equilibrium in the comoving frame is assumed.
Diffusion of nuclear species relative to the average flow velocity of the fluid is neglected.
Recognizing that neutrinos play a role in systems in which relativity and nuclear reactions are both important, the presence of neutrino-related source terms is noted, without giving detailed accounts of neutrino transport (e.g. \cite{Cardall2012Conservative-31,Cardall2013Conservative-31}) or interactions (e.g. \cite{Burrows2006Neutrino-opacit}).
Units in which the speed of light $c = 1$ and the reduced Planck constant $\hbar = 1$ are used throughout.
Latin indices near the beginning of the alphabet ($a, b, \dots$) denote particle species.
In the relativistic treatment, greek indices denote spacetime components, latin indices near the middle of the alphabet ($i, j, \dots$) denote spatial components, and the metric signature $(-+++)$ is employed.

\section{\label{sec:Newtonian} Newtonian Formulation}

A conservative formulation of perfect Newtonian fluid dynamics comprises balance equations for mass density $D$, momentum density $\mathbf{S}$, and internal + kinetic energy density $E$, in each case relating the time derivative of a volume density to the divergence of a flux:
\begin{eqnarray}
\frac{\partial D}{\partial t} + \bm{\nabla} \cdot \mathbf{F}_D &=& 0, \label{eq:Newtonian_D} \\
\frac{\partial\mathbf{S}}{\partial t} +  \bm{\nabla} \cdot \mathbf{F}_\mathbf{S} &=& \mathbf{Q}_\mathbf{S}, \label{eq:Newtonian_S} \\
\frac{\partial E}{\partial t} +  \bm{\nabla} \cdot \mathbf{F}_E &=& Q_E. \label{eq:Newtonian_E}
\end{eqnarray}
In terms of primitive variables---the mass density $\rho$, three-velocity $\mathbf{v}$, and internal energy density $e$---the conserved densities are
\begin{eqnarray}
D &=& \rho, \label{eq:D_Density_Newtonian} \\
\mathbf{S} &=& \rho \,\mathbf{v}, \\
E &=& e + \frac{1}{2}\, \rho \left(\mathbf{v} \cdot \mathbf{v}\right), \label{eq:E_Density_Newtonian}
\end{eqnarray}
and the fluxes are
\begin{eqnarray}
\mathbf{F}_D &=& \rho \,\mathbf{v}, \label{eq:D_Flux_Newtonian} \\
\mathbf{F}_\mathbf{S} &=& \rho \left(\mathbf{v} \otimes \mathbf{v}\right) + p\, \mathbf{I}, \\
\mathbf{F}_E &=& \left[ e + \frac{1}{2}\, \rho \left(\mathbf{v} \cdot \mathbf{v}\right) + p \right]\mathbf{v}, \label{eq:E_Flux_Newtonian}
\end{eqnarray}
in which $\mathbf{I}$ is the identity tensor, and the pressure $p$ is given by an equation of state, which depends on the system; absent composition variations, it can be as simple as $p = p(\rho,e)$, or even $p = p(\rho)$.

Allowing for gravity and neutrino interactions, and separating the contribution of nuclear binding to the internal energy density, the sources are
\begin{eqnarray}
\mathbf{Q}_\mathbf{S} &=& -\rho\, \bm{\nabla} \Phi + \mathbf{A}_{\Sigma\nu}, \label{eq:S_Source_Newtonian} \\
Q_E &=& -\rho\,\mathbf{v} \cdot \bm{\nabla} \Phi + Q_{\Sigma\nu} + Q_{\Delta M}. \label{eq:E_Source_Newtonian}
\end{eqnarray}
The gravitational potential $\Phi$ satisfies the Poisson equation $\nabla^2 \Phi = 4\pi G_N\rho$, with Newton's constant $G_N$.
The neutrino sources $\mathbf{A}_{\Sigma\nu}$ and $Q_{\Sigma\nu}$ include contributions from all neutrino species.
The distribution functions $f_{\nu_a}(t, \mathbf{x}, \mathbf{p})$ of each neutrino type $\nu_a$ satisfy the Boltzmann equation 
\begin{equation}
\frac{df_{\nu_a}}{dt} = \frac{\partial f_{\nu_a}}{ \partial t} + \hat\mathbf{n} \cdot \bm{\nabla} f_{\nu_a} = R_{\nu_a},
\end{equation}
where $\hat\mathbf{n} = \mathbf{p} / \epsilon$ is the neutrino momentum direction and $\epsilon = |\mathbf{p}|$ is the neutrino energy, both measured by an Eulerian observer; and $R_{\nu_a}$ is the collision integral. 
The momentum gained by the fluid is that lost by neutrinos:
\begin{eqnarray}
\mathbf{A}_{\Sigma\nu} &=& - \frac{d}{dt} \left( \sum_{\nu_a} \int \frac{d\mathbf{p}}{(2\pi)^3} \, \mathbf{p} \, f_{\nu_a} \right) \\
&=& - \sum_{\nu_a} \int \frac{d\mathbf{p}}{(2\pi)^3} \, \mathbf{p} \, R_{\nu_a},
\end{eqnarray}
and similarly
\begin{equation}
Q_{\Sigma\nu} = -\sum_{\nu_a} \int \frac{d\mathbf{p}}{(2\pi)^3} \, \epsilon \, R_{\nu_a}
\end{equation}
for the neutrino energy source.
The neutrino heating $Q_{\Sigma\nu}$ and nuclear energy generation $Q_{\Delta M}$
both arise from the first law of thermodynamics applied to a perfect fluid mass element,
\begin{equation}
\frac{d}{dt}\left(\frac{e + b}{\rho} \right) = - \,p\, \frac{d}{dt}\left(\frac{1}{\rho} \right) + \frac{q_{\Sigma\nu}}{\rho},
\label{eq:FirstLaw}
\end{equation}
used in the derivation of Eq.~(\ref{eq:Newtonian_E}).
Here $b$ is the energy density due to nuclear binding, with $e$ comprising all other contributions to internal energy density.
Moreover $q_{\Sigma\nu}$ is the energy density absorption rate experienced by a Lagrangian mass element; taking into account the Newtonian Doppler shift, 
\begin{eqnarray}
q_{\Sigma\nu} &=& - \sum_{\nu_a} \int \frac{d\mathbf{p}}{(2\pi)^3} \, \epsilon\, ( 1 - \mathbf{v} \cdot \hat\mathbf{n} ) \, R_{\nu_a} \\
&=& Q_{\Sigma\nu} - \mathbf{v} \cdot \mathbf{A}_{\Sigma\nu}.
\end{eqnarray} 
(Note the cancellation of the $\mathbf{v} \cdot \mathbf{A}_{\Sigma\nu}$ term when arriving at Eq.~(\ref{eq:Newtonian_E}) with the help of Eq.~(\ref{eq:Newtonian_S})).
Finally, the separation of the energy density $b$ due to nuclear binding from the rest of the internal energy density in Eq.~(\ref{eq:FirstLaw}) leads to
\begin{equation}
Q_{\Delta M} =  - \rho\,\frac{d}{dt}\left( \frac{b}{\rho} \right) \label{eq:E_Source_Nuclear}
\end{equation}
for the energy source term due to nuclear burning.

The evolution of the nuclear species must be addressed in order to flesh out the nuclear energy generation source term $Q_{\Delta M}$ of Eq.~(\ref{eq:E_Source_Nuclear}).
The number density $n_a$ of each nuclear species $a$ satisfies a balance equation of the form
\begin{equation}
\frac{\partial n_a}{\partial t} + \boldsymbol\nabla \cdot \mathbf{F}_{n_a} = R_a.\label{eq:Newtonian_N_a}
\end{equation}
The traditional primitive composition variable is the abundance $Y_a$, defined 
such that
\begin{eqnarray}
n_a &=& \frac{\rho\, Y_a}{m_u}, \label{eq:Y_a} \\
\mathbf{F}_{n_a} &=& \frac{\rho\, Y_a}{m_u}\, \mathbf{v}, \label{eq:N_a_Flux_Newtonian}
\end{eqnarray}
in which $m_u$ is the atomic mass unit, and diffusion of species relative to the average flow velocity $\mathbf{v}$ of the fluid is neglected.
The source $R_a = R_a \left( \rho, T, \left\{Y_b\right\}\right)$ is the net particle production rate per unit volume, with $T$ being the local fluid temperature.
The utility of the abundance $Y_a$ as a composition variable is that, thanks to mass continuity as expressed in Eqs.~(\ref{eq:Newtonian_D}), (\ref{eq:D_Density_Newtonian}), (\ref{eq:D_Flux_Newtonian}),
its Lagrangian derivative 
\begin{equation}
\frac{dY_a}{dt} = \frac{m_u }{\rho} \, R_a\label{eq:dYdt}
\end{equation}
eliminates the effects of compression, changing only due to reactions (in the absence of diffusion). 

Only energy differences matter in the Newtonian limit, so that the binding energy density $b$ can be measured relative to an arbitrary fixed reference point.
While there is freedom in the choice of reference point, that reference point must indeed be fixed.
In the presence of weak interactions, as contemplated here, use of the absolute binding energy of a nucleus or neutral atom---defined as the difference between its mass and the sum of the masses of its free constituent particles---is not suitable, because the total proton and neutron numbers in a mass element are not separately conserved.
Instead, a reference point that depends only on the total nucleon number is required.   
One possible choice is the mass excess $(\Delta M)_a$, the binding energy relative to that of $^{12}$C: 
\begin{equation}
(\Delta M)_a = M_a - m_u \,A_a, \label{eq:MassExcess}
\end{equation}
where $M_a$ is the mass of a single nucleus of species $a$ and $A_a$ is its nucleon number.
Then the energy density due to nuclear binding is
\begin{equation}
b = \sum_a (\Delta M)_a n_a, \label{eq:RelativeBinding}
\end{equation}
so that, using Eq.~(\ref{eq:Y_a}),
\begin{equation}
- \rho\,\frac{d}{dt}\left( \frac{b}{\rho} \right) = - \frac{\rho}{m_u} \sum_a (\Delta M)_a \frac{dY_a}{dt}.
\end{equation}
Thus Eqs.~(\ref{eq:E_Source_Nuclear}) and (\ref{eq:dYdt}) yield 
\begin{equation}
Q_{\Delta M} = - \sum_a (\Delta M)_a R_a, \label{eq:EnergyGeneration_Delta_M}
\end{equation}
completing the description of the energy equation source term when the nuclear binding energy is separated from the rest of the internal energy (and diffusion is neglected).

As a segue into the relativistic formulation in the next section, consider a retrofit of the Newtonian perspective effected by relativistic insight into the nature of the mass density in the presence of composition changes.
Nuclear reactions do not change the total baryon number density
\begin{equation}
n = \sum_a A_a n_a, \label{eq:BaryonNumberDensity}
\end{equation} 
so $n$ obeys the conservation law
\begin{equation}
\frac{\partial n}{\partial t} + \bm{\nabla}\cdot (n \mathbf{v}) = 0. \label{eq:Newtonian_N}
\end{equation}
This is only consistent with (what has improvidently been dubbed `mass conservation' in) Eqs.~(\ref{eq:Newtonian_D}), (\ref{eq:D_Density_Newtonian}), (\ref{eq:D_Flux_Newtonian}) if $\rho$ is a \textit{constant} multiple of $n$, for instance
\begin{equation}
\rho \equiv m_u n. \label{eq:rho}
\end{equation}
This choice makes $\rho$ numerically close to the true baryon mass density
\begin{equation}
\rho_m = \sum_a M_a n_a \equiv m\,n, \label{eq:rho_m}
\end{equation}
in which a \textit{mutable} average baryon mass
\begin{equation}
m = \frac{1}{n} \sum_a M_a n_a = \sum_a M_a Y_a \label{eq:AverageBaryonMass}
\end{equation}
has been defined \footnote{When $M_a$ are the measured masses of neutral atoms, the `baryon mass density' $\rho_m$ includes the mass of electrons as well (modulo ionization energy, which however is not normally regarded as significant compared with nuclear binding energy). 
At temperatures high enough for positrons to be present, this convention implies that $\rho_m$ includes the mass $m_e ( n_{e^-} - n_{e^+} )$ of the net number of electrons balancing the total proton charge.
Then the internal energy $e$ must include, in addition to thermal energy, the mass density $2\, m_e n_{e^+}$ of $e^+ e^-$ pairs (cf. the `$e_\mathrm{th}$' in Appendix~A of Ref.~\cite{Bruenn2016The-Development}); but this is a relatively small contribution when pairs are abundant.}.
The second equality follows from Eqs.~(\ref{eq:Y_a}) and (\ref{eq:rho}),
according to which the abundance
\begin{equation}
Y_a = \frac{n_a}{n}  \label{eq:Y_a_Strict}
\end{equation}
is strictly the number density of species $a$ relative to the total baryon number density.
Note that Eqs.~(\ref{eq:rho})-(\ref{eq:rho_m}) and (\ref{eq:MassExcess})-(\ref{eq:RelativeBinding}) yield
\begin{equation}
\rho_m = \rho + b.
\end{equation}
In practice the relative binding energy density $b$ is only crucial in comparison with the internal energy density $e$, and the distinction between $\rho$ and $\rho_m$ is often swept under the rug.

However, it is worth keeping in mind---both for conceptual clarity, and for the relativistic derivation in the next section---that the true baryon mass density of Eq.~(\ref{eq:rho_m}) obeys a balance equation rather than a strict conservation law:
\begin{eqnarray}
\frac{\partial \rho_m}{\partial t} + \bm{\nabla}\cdot (\rho_m \mathbf{v}) &=& \frac{\partial( m\, n)}{\partial t} + \bm{\nabla}\cdot (m \, n \,\mathbf{v}) \nonumber \\
&=& n\frac{dm}{dt}, \label{eq:Newtonian_rho_m}
\end{eqnarray}
with the second line following from Eq.~(\ref{eq:Newtonian_N}), and
\begin{equation}
n \frac{dm}{dt} = \sum_a M_a R_a \label{eq:dmdt}
\end{equation}
from Eqs.~(\ref{eq:dYdt}) and (\ref{eq:AverageBaryonMass}). 
Using the identity 
\begin{eqnarray}
0 &=& n\, \frac{d}{dt}\left(\frac{1}{n} \sum_a A_a n_a \right) \nonumber \\
&=& n \sum_a A_a \frac{dY_a}{dt} = \sum_a A_a R_a
\end{eqnarray} 
following from Eqs.~(\ref{eq:dYdt}), (\ref{eq:BaryonNumberDensity}) and (\ref{eq:Y_a_Strict}), subtracting the vanishing constant $0 = m_u \sum_a A_a R_a$ from Eq.~(\ref{eq:dmdt}) reveals the source
\begin{equation}
n \frac{dm}{dt} = \sum_a \left(\Delta M\right)_a R_a  = - \,Q_{\Delta M} \label{eq:dmdt_Delta_M}
\end{equation}
in Eq.~(\ref{eq:Newtonian_rho_m}) to be none other than the nuclear energy generation encountered in Eq.~(\ref{eq:EnergyGeneration_Delta_M}).

\section{Relativistic Formulation}
\label{sec:Relativistic}

In the relativistic case the balance equations governing a perfect fluid follow from the spacetime divergences of the baryon number flux vector $N^\mu$ and fluid stress-energy tensor $T^{\mu\nu}$:
\begin{eqnarray}
\nabla_\mu N^\mu &=& 0, \label{eq:Divergence_N} \\
\nabla_\mu T^{\mu\nu} &=& \left(Q_{\Sigma\nu} \right)^\nu, \label{eq:Divergence_T}
\end{eqnarray}
where $\nabla_\mu$ is the covariant derivative and 
\begin{equation}
\left(Q_{\Sigma\nu} \right)^\nu = - \nabla_\mu \left( T_{\Sigma\nu} \right)^{\mu\nu}
\end{equation}
is the four-momentum source due to neutrino interactions, arising from the divergence of the stress-energy tensor $\left( T_{\Sigma\nu} \right)^{\mu\nu}$ of all neutrino species (with apologies for the subscript $\Sigma\nu$, which stands for `all neutrino species' and is not a spacetime index).
The baryon number flux vector is 
\begin{equation}
N^\mu = n\, u^\mu, \label{eq:NumberFlux_N}
\end{equation}
where $n$ is the baryon number density in the comoving frame and $u^\mu$ is the fluid four-velocity.
The fluid stress-energy tensor is 
\begin{equation}
T^{\mu\nu} = (m\, n  + e + p) \, u^\mu u^\nu + p \, g^{\mu\nu}, \label{eq:T_perfect}
\end{equation}
in which $m$ is the average baryon mass discussed in the last two paragraphs of Sec.~\ref{sec:Newtonian}.
Note that $e$ does not include nuclear binding energy, which instead contributes to $m$.  

The $3+1$ formulation of general relativity is useful in bringing Eqs.~(\ref{eq:Divergence_N}) and (\ref{eq:Divergence_T}) closer to a form that is simulation-ready and more reminiscent of Newtonian fluid dynamics.
Central to this approach is a foliation of spacetime into spacelike slices $\Sigma_t$ labeled by time component $t$, the component $x^0$ of spacetime position $x^\mu$. 
The unit normal $n_\mu$ to a spacelike slice (a tensor not to be confused with the scalar baryon number density $n$) has components
\begin{eqnarray}
\left( n_\mu \right) &=& \left( -\alpha, 0, 0, 0\right), \label{eq:n_D} \\
\left( n^\mu \right) &=& \left( 1/\alpha, -\beta^i / \alpha \right)^T.
\end{eqnarray} 
Consider an infinitesimal perpendicular displacement (i.e., parallel to $n^\mu$) connecting slices $\Sigma_t$ and $\Sigma_{t+dt}$.
The lapse function $\alpha$ gives the proper time $\alpha\, dt$ separating the slices along this displacement.
If $x^i$ is the base of the displacement in $\Sigma_t$, then the shift vector $\beta^i$ gives the proper distance $\beta^i\, dt$ between the tip of the perpendicular displacement in $\Sigma_{t+dt}$ and coordinate position $x^i$ in $\Sigma_{t+dt}$.
The induced three-metric $\gamma_{ij}$ characterizes the geometry within a spacelike slice.
These geometric prescriptions are encapsulated in the spacetime line element
\begin{equation}
ds^2 = -\alpha^2\, dt^2 + \gamma_{ij} \left(dx^i + \beta^i\, dt \right) \left(dx^j + \beta^j\, dt \right),
\end{equation}
from which the components of the spacetime metric $g_{\mu\nu}$ can be read.
The determinant $g$ of the spacetime metric is related to the determinant $\gamma$ of the induced three-metric by $\sqrt{-g} = \alpha \sqrt{\gamma}$.
The tensor
\begin{equation}
\gamma_{\mu\nu} = g_{\mu\nu} + n_\mu n_\nu
\end{equation}
projects components tangent to a spacelike slice.
 
Also useful for making contact with the Newtonian fluid dynamics equations is the three-velocity four-vector $v^\mu$, defined by decomposing the fluid four-velocity $u^\mu$ into parts normal and tangent to the spacelike slice:
\begin{equation}
u^\mu = \Lambda \left( n^\mu + v^\mu \right), \label{eq:u_n_v}
\end{equation}
with $n_\mu v^\mu = 0$.
Together with Eq.~(\ref{eq:n_D}), this requires that component $v^0 = 0$.
The interpretation of $v^\mu$ as a three-velocity is confirmed by squaring Eq.~(\ref{eq:u_n_v}), which yields 
\begin{equation}
\Lambda = \left(1 - v_\mu v^\mu \right)^{-1/2} = \left(1 - v_i v^i \right)^{-1/2}, \label{eq:LorentzFactor}
\end{equation}
showing that $\Lambda$ can be regarded as the Lorentz factor connecting an Eulerian observer with four-velocity $n^\mu$ to a Lagrangian observer with four-velocity $u^\mu$.

Using the $3+1$ formalism and the fluid three-velocity $v^\mu$, Eq.~(\ref{eq:Divergence_N}) for baryon number becomes
\begin{equation}
\frac{1}{\alpha\sqrt{\gamma}} \frac{\partial }{\partial t} \left(\sqrt{\gamma}\, N\right) + \frac{1}{\alpha\sqrt{\gamma}} \frac{\partial }{\partial x^i} \left[ \sqrt{\gamma} \left(F_N\right)^i\right] = 0, \label{eq:Relativistic_N}
\end{equation}
with the conserved density and flux
\begin{eqnarray}
N &=& \Lambda \, n, \\
(F_N)^i &=& \Lambda \, n \left( \alpha\, v^i - \beta^i \right).
\end{eqnarray}
This agrees with Eq.~(\ref{eq:Newtonian_D}) for Newtonian mass conservation if composition changes are ignored, so that the average baryon mass in $\rho = m\, n$ can pass through derivatives in Eq.~(\ref{eq:Relativistic_N}); if bulk fluid speeds are much smaller than the speed of light ($\Lambda \rightarrow 1$); and if spacetime curvature is unimportant ($\alpha \rightarrow 1$, $\beta^i \rightarrow 0$, and $\sqrt{\gamma}$ is independent of $t$ and reflects only the use of flat-space curvilinear coordinates).

The momentum equation corresponding to Eq.~(\ref{eq:Newtonian_S}) follows from the spacelike projection
\begin{equation}
\gamma_{j\nu} \, \nabla_\mu T^{\mu\nu} = \left(A_{\Sigma\nu} \right)_j
\end{equation}
of Eq.~(\ref{eq:Divergence_T}), where $\left(A_{\Sigma\nu} \right)_j = \gamma_{j\mu} \, \left(Q_{\Sigma\nu} \right)^\mu$ is the momentum source due to neutrinos.
For generic $T^{\mu\nu}$ decomposed as
\begin{eqnarray}
T^{\mu\nu} &=& G\, n^\mu n^\nu + S^\mu \, n^\nu + S^\nu \, n^\mu + P^{\mu\nu}, \\
0 &=& n_\mu \, S^\mu, \\
0 &=& n_\mu \, P^{\mu\nu} = n_\nu\, P^{\mu\nu},
\end{eqnarray}
this is worked out in (for example) Appendix~A of Cardall et al. (2013) as
\begin{equation}
\frac{1}{\alpha\sqrt{\gamma}} \frac{\partial }{\partial t} \left(\sqrt{\gamma}\, S_j\right) + \frac{1}{\alpha\sqrt{\gamma}} \frac{\partial }{\partial x^i} \left[\sqrt{\gamma}\, {(F_S)_j}^i \right] = (Q_S)_j, \label{eq:Relativistic_S}
\end{equation}
with 
\begin{eqnarray}
{(F_S)_j}^i &=& \alpha \, {P_j}^i - S_j\,\beta^i \,, \\
(Q_S)_j &=& - \frac{G}{\alpha} \frac{\partial\alpha}{\partial x^j} + \frac{S_i}{\alpha}\frac{\partial\beta^i}{\partial x^j} + \frac{P^{ik}}{2} \frac{\partial\gamma_{ik}}{\partial x^j} \nonumber \\
& & + \left(A_{\Sigma\nu} \right)_j
\end{eqnarray}
(see also for instance Refs.~\cite{Shibata2005Magnetohydrodyn} and \cite{Del-Zanna2007ECHO:-a-Euleria} for the metric source terms in this form).
For the perfect fluid stress-energy tensor of Eq.~(\ref{eq:T_perfect}),
\begin{eqnarray}
G &=& n_\mu n_\nu \, T^{\mu\nu} = \Lambda^2 (m\, n  + e + p) \, - \, p, \label{eq:Projection_G} \\
S_j &=& - \gamma_{j \mu} n_\nu \, T^{\mu\nu} = \Lambda^2 (m\, n  + e + p)\, v_j, \label{eq:Projection_S} \\
{P_j}^i &=& \gamma_{j \mu} {\gamma^i}_{\nu} \, T^{\mu\nu} = \Lambda^2 (m\, n  + e + p)\, v_j v^i \nonumber \\
& & + \, p\, {\delta_j}^i.
\end{eqnarray}
The flux can be also be expressed as 
\begin{equation}
{(F_S)_j}^i = \Lambda^2 (m\, n  + e + p) \, v_j \left(\alpha\, v^i - \beta^i \right) + \alpha \, p\, {\delta_j}^i.
\end{equation}
Agreement with Newtonian momentum conservation in Sec.~\ref{sec:Newtonian} requires not only slow bulk fluid speeds ($\Lambda \rightarrow 1$), but also microscopic particle speeds much less than the speed of light, so that $e \ll m\,n$ and $p \ll m\, n$.
As for the metric functions in the Newtonian limit, once again $\alpha \rightarrow 1$, $\beta^i \rightarrow 0$, and $\gamma_{ij}$ represents only flat-space curvilinear coordinates, except that retention of the leading term of the Newtonian limit $-g_{00} \rightarrow \alpha^2 \rightarrow (1 + 2\Phi)$ in $\partial \alpha / \partial x^i$ gives the Newtonian gravitational acceleration.

The timelike projection
\begin{equation}
- n_\nu \, \nabla_\mu T^{\mu\nu} =  Q_{\Sigma\nu} \label{eq:Divergence_T_n}
\end{equation}
of Eq.~(\ref{eq:Divergence_T}), where $Q_{\Sigma\nu} = -n_\mu \, \left(Q_{\Sigma\nu} \right)^\mu$ is the energy source due to neutrinos, provides an equation for total relativistic fluid energy.
Again this is worked out in (for example) Appendix~A of Cardall et al. (2013), as
\begin{equation}
\frac{1}{\alpha\sqrt{\gamma}} \frac{\partial }{\partial t} \left(\sqrt{\gamma}\, G \right) + \frac{1}{\alpha\sqrt{\gamma}} \frac{\partial }{\partial x^i} \left[\sqrt{\gamma}\, (F_G)^i \right] = Q_G, \label{eq:Relativistic_G}
\end{equation}
with 
\begin{eqnarray}
(F_G)^i &=& \alpha \, S^i - \beta^i \,G, \\
Q_G &=& - \frac{S^i}{\alpha} \frac{\partial\alpha}{\partial x^i}  + P^{ij} K_{ij} + Q_{\Sigma\nu},
\end{eqnarray}
in which the extrinsic curvature tensor $K_{ij}$ describes the embedding of the spacelike slices in spacetime (see also for instance Refs.~\cite{Shibata2005Magnetohydrodyn} and \cite{Del-Zanna2007ECHO:-a-Euleria} for the metric source terms in this form).
The flux can be rewritten as 
\begin{equation}
(F_G)^i = \Lambda^2 (m\, n  + e + p) \left( \alpha \, v^i - \beta^i \right) \, + \, \beta^i \, p
\end{equation}
in light of Eqs.~(\ref{eq:Projection_G}) and (\ref{eq:Projection_S}).

Equation~(\ref{eq:Relativistic_G}) may at first seem puzzling when its Newtonian limit is pondered.
If the limit $\Lambda \rightarrow 1$ is not taken everywhere, but instead terms of $\mathcal{O}(v^2)$ are kept when they multiply $m\, n$ in order to get the kinetic energy density,
then from Eqs.~(\ref{eq:Projection_G}) and (\ref{eq:LorentzFactor}),
\begin{equation}
G \rightarrow m \, n \left( 1 + v_i v^i \right) \, + \, e. \label{eq:G_Limit}
\end{equation}
Aside from the baryon rest energy density $m\, n$, there is a factor of 2 difference in the $v^2$ term relative to Eq.~(\ref{eq:E_Density_Newtonian}).

However, an appropriate combination of the energy Eq.~(\ref{eq:Relativistic_G}) with baryon conservation yields an energy equation that \textit{does} reduce to Eq.~(\ref{eq:Newtonian_E}), and as a natural by-product automatically separates out the energy generation due to nuclear reactions.
Multiplying Eq.~(\ref{eq:Divergence_N}) by average baryon mass $m$ and using the derivative product rule yields an additional term involving the spacetime gradient of $m$:
\begin{equation}
\nabla_\mu \left( m\, N^\mu \right) - N^\mu \,\nabla_\mu m = 0, \label{eq:Divergence_D}
\end{equation}
or
\begin{equation}
\frac{1}{\alpha\sqrt{\gamma}} \frac{\partial }{\partial t} \left(\sqrt{\gamma}\, D\right) + \frac{1}{\alpha\sqrt{\gamma}} \frac{\partial }{\partial x^i} \left[ \sqrt{\gamma} \left( F_D \right)^i\right] = Q_D, \label{eq:Relativistic_D}
\end{equation}
where
\begin{eqnarray}
D &=& \Lambda\, m\, n, \\
\left(F_D \right)^i &=& \Lambda \, m\, n \left(\alpha\, v^i - \beta^i \right), \\
Q_D &=& n \, u^\mu \frac{\partial m}{\partial x^\mu} \equiv n \, \dot m
\label{eq:D_Source_Relativistic}
\end{eqnarray}
(compare Eq.~(\ref{eq:Newtonian_rho_m}) in the Newtonian case).
Subtracting Eq.~(\ref{eq:Relativistic_D}) from Eq.~(\ref{eq:Relativistic_G}) results in the analogue of the Newtonian energy Eq.~(\ref{eq:Newtonian_E}):
\begin{equation}
\frac{1}{\alpha\sqrt{\gamma}} \frac{\partial }{\partial t} \left(\sqrt{\gamma}\, E \right) + \frac{1}{\alpha\sqrt{\gamma}} \frac{\partial }{\partial x^i} \left[\sqrt{\gamma}\, (F_E)^i \right] = Q_E, \label{eq:Relativistic_E}
\end{equation}
with
\begin{eqnarray}
E &=&  \Lambda^2 ( e + p) \, - \, p  \, + \, \Lambda \left( \Lambda - 1 \right) m\, n, \label{eq:E_Density_Relativistic} \\
(F_E)^i &=& \left[ \Lambda^2 ( e + p) \,+\, \Lambda \left( \Lambda - 1 \right) m\, n \right]\left( \alpha \, v^i - \beta^i \right) \nonumber\\
& &+ \, \beta^i \, p, \label{eq:E_Flux_Relativistic}
 \\
Q_E &=& - \frac{S^i}{\alpha} \frac{\partial\alpha}{\partial x^i}  + P^{ij} K_{ij} + Q_{\Sigma\nu} - n \, \dot m. \label{eq:E_Source_Relativistic}
\end{eqnarray}
For slow bulk fluid and microscopic particle speeds ($\Lambda \rightarrow 1$; $e, \; p,\;  m\, n\, v^2 \ll m\, n$) and vanishing spacetime curvature ($\alpha \rightarrow 1, \beta^i \rightarrow 0, K_{ij} \rightarrow 0$, $\gamma_{ij}$ represents only flat-space curvilinear coordinates) except for $\partial \alpha / \partial x^i \rightarrow \partial \Phi / \partial x^i$, 
it is apparent that Eqs.~(\ref{eq:E_Density_Relativistic})-(\ref{eq:E_Source_Relativistic}) reduce to the Newtonian Eqs.~(\ref{eq:E_Density_Newtonian}), (\ref{eq:E_Flux_Newtonian}), and (\ref{eq:E_Source_Newtonian}). 
In particular, 
\begin{equation}
\Lambda \left( \Lambda - 1 \right) m\, n \rightarrow \frac{m\, n}{2}\, v_i v^i
\end{equation}
reduces to the Newtonian kinetic energy with its factor of $1/2$ as expected and desired.

It remains to confirm more explicitly that the source term $- n \, \dot m$ in Eq.~(\ref{eq:E_Source_Relativistic}) agrees with the nuclear energy generation source $Q_{\Delta M}$ of Eq.~(\ref{eq:EnergyGeneration_Delta_M}) in Sec.~\ref{sec:Newtonian}.
First note that the number flux of each species $a$ satisfies a divergence equation similar to Eq.~(\ref{eq:Divergence_N}), but with a source, the same number production rate $R_a$ as that in Eq.~(\ref{eq:Newtonian_N_a}):
\begin{equation}
\nabla_\mu \left(N_a\right)^\mu = R_a, \label{eq:Divergence_N_a}
\end{equation}
with 
\begin{equation}
(N_a)^\mu = n_a u^\mu = n\, Y_a u^\mu, \label{eq:NumberFlux_N_a}
\end{equation}
where $n_a$ is the number density of species $a$ in the comoving frame and $Y_a = n_a / n$ is the abundance.
In terms of 3+1 coordinates,
\begin{equation}
\frac{1}{\alpha\sqrt{\gamma}} \frac{\partial }{\partial t} \left(\sqrt{\gamma}\, N_a \right) + \frac{1}{\alpha\sqrt{\gamma}} \frac{\partial }{\partial x^i} \left[ \sqrt{\gamma} \left(F_{N_a}\right)^i\right] = R_a, \label{eq:Relativistic_N_a}
\end{equation}
with the conserved density and flux
\begin{eqnarray}
N_a &=& \Lambda \, n_a = \Lambda\, n\, Y_a, \\
(F_{N_a})^i &=& \Lambda \, n\, Y_a \left( \alpha\, v^i - \beta^i \right).
\end{eqnarray}
Alternatively, using the last expression in Eq.~(\ref{eq:NumberFlux_N_a}) in Eq.~(\ref{eq:Divergence_N_a}) and making use of baryon conservation in Eqs.~(\ref{eq:Divergence_N}) and (\ref{eq:NumberFlux_N}), one finds
\begin{equation}
n \,\dot{Y_a} \equiv n\, u^\mu \,\frac{\partial Y_a}{\partial x^\mu} = R_a,
\end{equation}
the relativistic analogue of Eq.~(\ref{eq:dYdt}).
The average baryon mass $m$ has the same definition as in Eq.~(\ref{eq:AverageBaryonMass}) in Sec.~\ref{sec:Newtonian}.
Moreover, $\dot m = u^\mu\, \partial m / \partial x^\mu$---the directional derivative along fluid four-velocity $u^\mu$---is the derivative of $m$ with respect to proper time, that is, the time derivative in the comoving frame.
This is the relativistic analogue of the Lagrangian time derivative $d/dt$ in Sec.~\ref{sec:Newtonian}.
With $dm/dt \rightarrow \dot m$ and $dY_a / dt \rightarrow \dot{Y_a}$, arguments parallel to those in the last paragraph of Sec.~\ref{sec:Newtonian} hold (including $\sum_a A_a R_a = 0$), with the result that
\begin{equation}
- n \,\dot m =  - \sum_a (\Delta M)_a \, R_a  =  Q_{\Delta M} 
\end{equation}
in the relativistic case as well.

\section{Conclusion}
\label{sec:Conclusion}

The purpose of this paper is to derive a relativistic fluid energy equation in conservative form 
in which the nuclear energy generation is explicitly separated, the baryon rest mass density is not present in the energy density, and whose Newtonian limit (including the correct factor of 2 in the kinetic energy density) matches Eq.~(\ref{eq:Newtonian_E}).
This result is achieved very simply, by subtracting the balance equation for mass density, Eq.~(\ref{eq:Relativistic_D}), from the balance equation for total energy, Eq.~(\ref{eq:Relativistic_G}), to yield Eq.~(\ref{eq:Relativistic_E}).

Given that---in mathematical terms---this result is arrived at almost trivially, it is worth pausing to ask why it does not seem to have been previously noted in the literature on relativistic fluid dynamics. Part of the answer may be that the deployment of nuclear reaction networks within numerical relativity codes is not yet widespread, so that the issue simply may have not been noticed.
But another, perhaps more important, reason may be the manner in which the relationship between relativistic and Newtonian fluid dynamics is often conceptualized.

(In fact, the energy variable here denoted $E = G - D$ has been widely (though not universally) used at least since the seminal introduction of conservative formulations to relativistic numerical hydrodynamics in spherical symmetry \cite{Marti1991Numerical-relat} and in three spatial dimensions \cite{Banyuls1997Numerical-3--1-}.
The motivation universally cited, when one is given, is numerical: use of the relativistic total fluid energy density, with a strongly dominant baryon rest mass density, can make it numerically challenging to resolve relatively small changes in the internal energy density.
However, the presence of a source term in the continuity equation for \textit{mass} density (as opposed to \textit{baryon number} density) due to nuclear composition changes seems to have gone unremarked in the literature.
And perhaps surprisingly, even the fact that the relativistic $E = G - D$ reduces to the Newtonian $e + \rho\, v^2/2$, with the correct factor of $1/2$ in the kinetic energy---unlike $G$ itself---also seems not to have been explicitly noted.)

Consider for a moment the Newtonian perspective, applied to mass elements, on its own terms.
Equation~(\ref{eq:Newtonian_D}) represents a first independent Newtonian principle: conservation of mass.
Newton's second law, a second independent principle, is embodied by Eq.~(\ref{eq:Newtonian_S}).
However, Eq.~(\ref{eq:Newtonian_E}) typically would not be regarded as an independent principle.
Instead, the first law of thermodynamics, expressed in Lagrangian form appropriate for mass elements, Eq.~(\ref{eq:FirstLaw}), would be regarded as a third independent principle, with Eq.~(\ref{eq:Newtonian_E})---conservation of internal + kinetic energy---following as a derived consequence of combining Eq.~(\ref{eq:FirstLaw}) with Eqs.~(\ref{eq:Newtonian_D}) and (\ref{eq:Newtonian_S}).

Contrast this with the relativistic perspective on fluids, which starts conceptually from different principles.
Mass, recognized as just another form of energy, is not conserved. 
However, there is conservation of baryon number---or for present purposes, conservation of nucleons---a principle expressed in Eq.~(\ref{eq:Relativistic_N}).
The other major independent principle is covariant local conservation of energy-momentum, expressed by the vanishing divergence of the fluid + neutrino stress-energy tensor.
With the unit normal $n^\mu$ to the spacelike slice regarded as the four-velocity of an Eulerian observer, the projections $\gamma_{j\nu} \nabla_\mu T^{\mu\nu} = \left(A_{\Sigma\nu} \right)_j$ and $- n_\nu \nabla_\mu T^{\mu\nu} = Q_{\Sigma\nu}$ respectively give Eulerian momentum and total fluid energy conservation, as expressed in Eqs.~(\ref{eq:Relativistic_S}) and (\ref{eq:Relativistic_G}).
(Projections orthogonal and parallel to a Lagrangian observer's four-velocity $u^\mu$ yield instead the the relativistic Euler equation for velocity and the first law of thermodynamics as applied to a perfect fluid.)
Conservation of internal + kinetic energy, Eq.~(\ref{eq:Relativistic_E}), is once again a derived consequence, this time a combination of Eq.~(\ref{eq:Relativistic_G}) and average baryon mass $m$ times Eq.~(\ref{eq:Relativistic_N}), that is, Eq.~(\ref{eq:Relativistic_D}).

\begin{figure}[t!]
\includegraphics[width=3.3in]{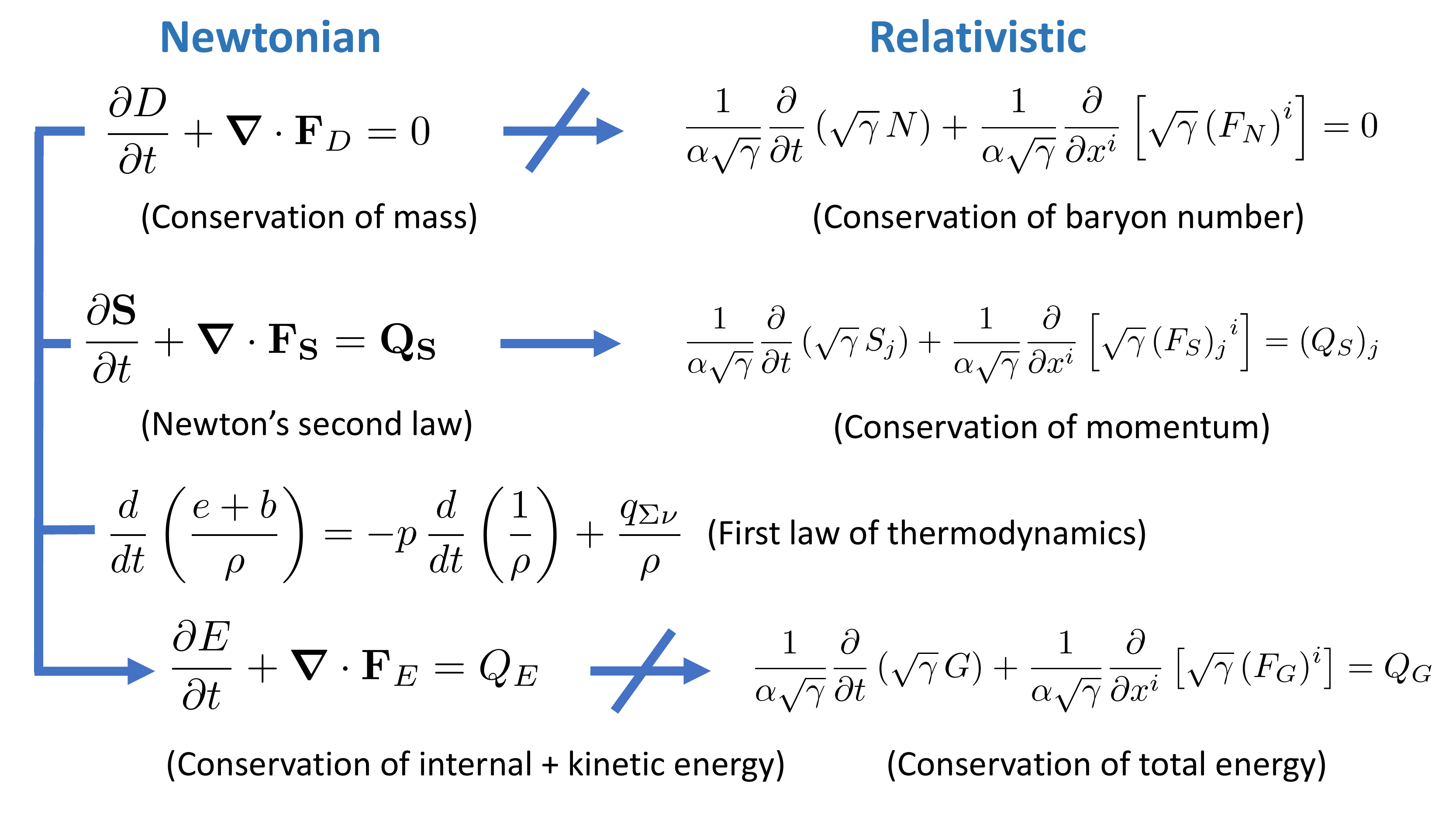}
\caption{A common but problematic correspondence between Newtonian and relativistic fluid dynamics.}
\label{fig:NewtonianLimitIncorrect}
\end{figure}

\begin{figure}[t!]
\includegraphics[width=3.3in]{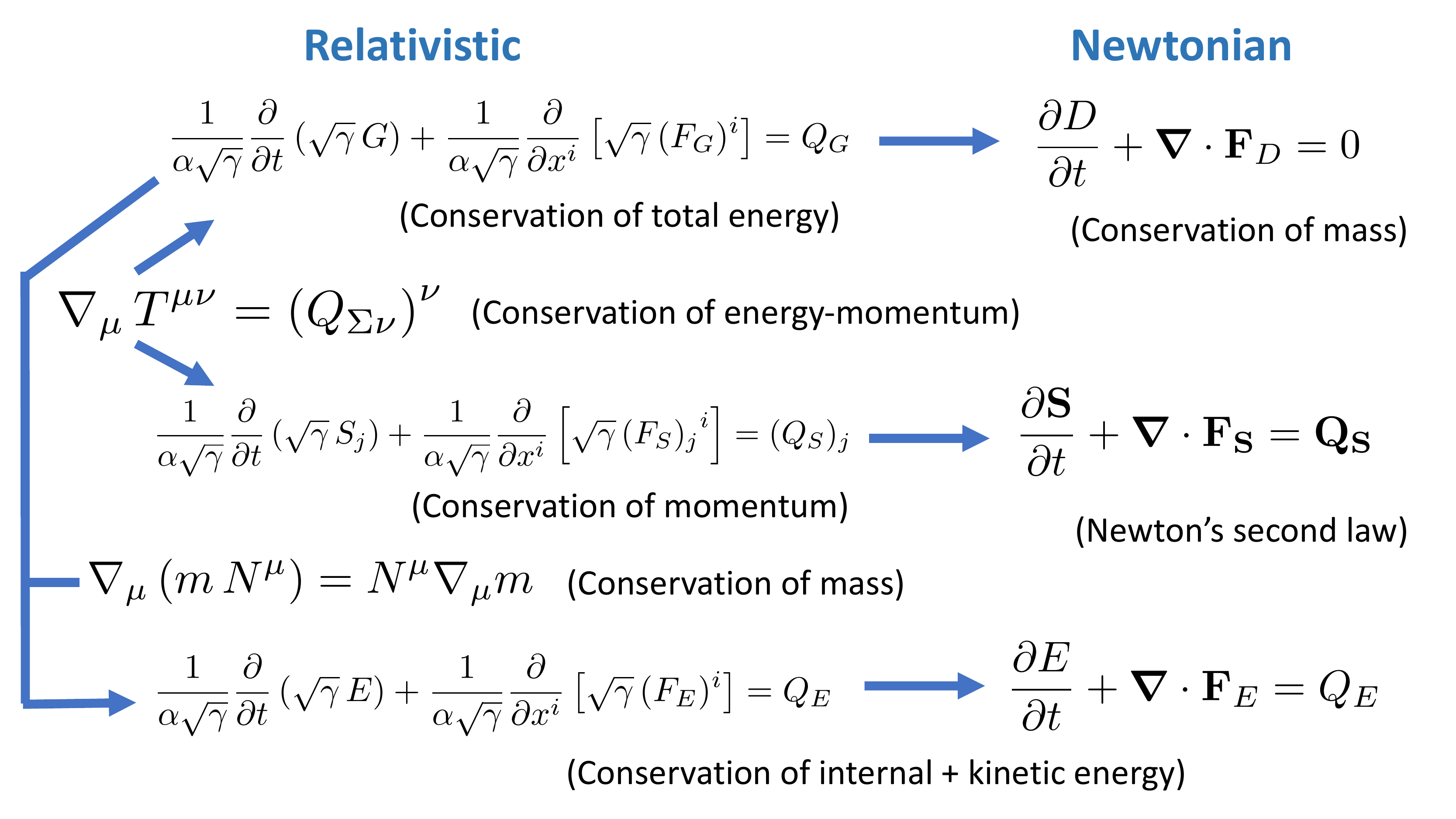}
\caption{A conceptually sound Newtonian limit of relativistic fluid dynamics.}
\label{fig:NewtonianLimitCorrect}
\end{figure}

There is an almost irresistible temptation, widely and understandably followed in the literature, to relate Newtonian and relativistic fluid dynamics as pictured in Fig.~\ref{fig:NewtonianLimitIncorrect}, which perhaps takes its cues from the Newtonian perspective: conservation of mass corresponds to baryon number conservation, Newton's second law corresponds to conservation of momentum, and conservation of `Newtonian total energy' (internal + kinetic) corresponds to relativistic total energy.
But in fact the first and third elements of this correspondence fail, due to the fundamental relativistic insight that mass is a form of energy. 
The phenomenon of nuclear energy generation cautions that conservation of baryon number is \textit{not} conservation of mass.
And relativistic total energy $G$ is fundamentally different from internal + kinetic energy $E$, both because of the presence of rest mass in $G$, and because of the low-velocity limit in Eq.~(\ref{eq:G_Limit}), which exhibits twice the Newtonian kinetic energy.

The more consistent relationship between relativistic and Newtonian fluid dynamics is that pictured in Fig.~\ref{fig:NewtonianLimitCorrect}. 
Newtonian conservation of mass is to be regarded not as an independent principle, on a par with conservation of baryon number; but as a derived consequence, the strictest limit of relativistic conservation of energy, Eq.~(\ref{eq:Relativistic_G}) with $\Lambda \rightarrow 1$,  $e, p \rightarrow 0$, and $\alpha \rightarrow 1$, $\beta^i, K_{ij} \rightarrow 0$, and $\gamma_{ij}$ representing only flat-space curvilinear coordinates 
\footnote{This is in line with the post-Newtonian formalism (e.g. \cite{Weinberg1972Gravitation-and,Misner1973Gravitation}), which regards the second and third terms of Eq.~(\ref{eq:G_Limit}) as the leading \textit{post}-Newtonian terms rather than part of the Newtonian limit.
In discussing the limit of $G$, Refs.~\cite{Landau1987Fluid-Mechanics,Gourgoulhon201231-Formalism-in} claim that the factor of $1/2$ does emerge, using arguments that, while perhaps related to what happens here with $E = G - D$, do not seem wholly satisfying---especially in light of the post-Newtonian perspective articulated in the  above references.}. 
Relativistic conservation of momentum does indeed correspond to Newton's second law. 
And conservation of relativistic internal + kinetic energy properly limits mathematically to conservation of Newtonian internal + kinetic energy---a derived quantity in both cases.

When adding nuclear energy generation to numerical relativity and fluid dynamics codes, the source term used in the Newtonian case admittedly could be blithely added to whatever energy equation is employed without significant numerical consequence; but conceptual clarity is worthwhile for its own sake.
True, the difference between baryon number density $n$ and its proxy in terms of mass density and atomic mass unit, $\rho / m_u$, is quantitatively insignificant in many or even most expressions. 
But the recurrent appearance of factors of $m_u$ is ugly \footnote{The alternative $\rho\, N_A$ in terms of Avogadro's number is an even less attractive alternative, as it binds one to cgs units.}, and a source of cognitive dissonance in the context of nucleosynthesis when one knows that nuclear energy generation is a nontrivial part of the problem.  
Moreover, when it comes to nucleosynthesis, the distinction is not trivial conceptually: when one does write down a relativistic balance equation for mass density, Eq.~(\ref{eq:Divergence_D}) or (\ref{eq:Relativistic_D}), the resulting source term, Eq.~(\ref{eq:D_Source_Relativistic}), is none other than nuclear energy generation itself!
Equation~(\ref{eq:Relativistic_E}) is a relativistic energy equation that separates nuclear energy generation in a conceptually coherent and self-consistent way, which limits to Eq.~(\ref{eq:Newtonian_E}) and maintains some features familiar from Newtonian numerical experience, but remains applicable to more extreme astrophysical conditions.

\appendix*
\section{}

While the focus of this paper is on the treatment of nuclear energy generation as a separate source term in an energy equation---the historically common practice when abundances are evolved with a nuclear network---a few comments on the alternative discussed in Sec.~\ref{sec:Introduction} are in order.
As mentioned there, nuclear binding energy can be included in the internal energy instead of as a separate source term, even when a nuclear network is employed; this would be accomplished with a binding energy density term $b$ in the internal energy, a sum over species like Eq.~(\ref{eq:RelativeBinding}).
This complicates the interface to the equation of state, in that the internal energy acquires a dependence on all abundances $\{Y_a\}$.
But one might want to do this if, for example, one is using (e.g. \cite{Harris2017Implications-fo,Bruenn2016The-Development}) a nuclear network at low density and a microphysical equation of state including nuclear forces at high density (including, for example, a phase transition to bulk nuclear matter):
if the high-density equation of state includes nuclear binding energy in its definition of internal energy, then one might wish to maintain consistency with that definition at all densities \footnote{As in the \textsc{Chimera} code \cite{Harris2017Implications-fo,Bruenn2016The-Development}; Raph Hix, private communication}.

In this case, instead of Eq.~(\ref{eq:T_perfect}), one would have the stress-energy tensor
\begin{equation}
T^{\mu\nu} = (m_B n  + \bar{e} + p) \, u^\mu u^\nu + p \, g^{\mu\nu}, \label{eq:T_perfect_2}
\end{equation}
in which the binding energy per nucleon taken account of in $\bar{e} = e + b$ is reckoned relative to a fixed reference baryon mass $m_B$.
(The atomic mass unit $m_u$ has been used as a reference in this paper, but this is not the only possible choice. 
For instance, the high-density equation of state of Lattimer and Swesty \cite{Lattimer1991A-generalized-e} measures baryon energies relative to the neutron mass $m_n$, so to match this convention one would take $m_B = m_n$ in Eq.~(\ref{eq:T_perfect_2}), and use binding energies relative to $A_a$ free neutrons in Eq.~(\ref{eq:RelativeBinding}).)
In a derivation paralleling that in Sec.~\ref{sec:Relativistic}, one arrives at Eq.~(\ref{eq:Relativistic_E}) for internal + kinetic energy, but with $e$ replaced by $\bar{e}$ in Eqs.~(\ref{eq:E_Density_Relativistic}) and (\ref{eq:E_Flux_Relativistic}) and elsewhere, and no $- n \,\dot{m}$ source term in Eq.~(\ref{eq:E_Source_Relativistic}).

\begin{acknowledgments}
Raph Hix provided valuable comments and corrections.
This work was supported by the U.S. Department of Energy, Office of Science, Office of Nuclear Physics under contract number DE-AC05-00OR22725.
\end{acknowledgments}

\def\apjs{Astrophys. J. Suppl. Ser. }
\def\aap{Astron. Astrophys. }


%

\end{document}